\documentstyle[aps,floats,epsfig,prl]{revtex}
\begin{document}
\draft
\title{Is a Trapped One-Dimensional Bose Gas a Luttinger Liquid?} 

\author{
  H. Monien, M. Linn, N. Elstner 
}
\address{
  Physikalisches Institut, 
  Universit\"at Bonn, \\
  Nu\ss allee 12,\\ 
  D-53115 Bonn,\\ 
  Germany 
}
\twocolumn[
\date{\today}
\maketitle
\widetext
\begin{abstract}
  \begin {center}
    \parbox{14cm}{ The low-energy fluctuations of a trapped, interacting quasi
      one-dimensional Bose gas are studied.  Our considerations apply to
      experiments with highly anisotropic traps. We show that under suitable
      experimental conditions the system can be described as a Luttinger
      liquid. This implies that the correlation function of the bosons decays
      algebraically preventing Bose-Einstein condensation. At significantly
      lower temperatures a finite size gap destroys the Luttinger liquid
      picture and Bose-Einstein condensation is again possible.  }
  \end{center}
\end{abstract}

\pacs{\hspace{1.9cm}PACS: 03.75.Fi, 05.30.jp, 32.80.Pj, 67.90.+z, 71.10.Pm}
]
\narrowtext
\vfill\eject

The experimental realization of Bose-Einstein condensation (BEC) in atomic
vapors of $^{87}$Rb \cite{Anderson95} and $^{23}$Na \cite{Davis95,Mewes96} has
attracted a lot of interest\cite{online}.  Recently, a highly anisotropic,
quasi one-dimensional trap has been designed \cite{MIT97}.  Up to now, the
possibility of BEC in one dimension has mainly been discussed for the
non-interacting Bose gas\cite{Bagnato91,Ketterle96}. The role of
dimensionality has been carefully examined for the {\em ideal} bose gas by van
Druten and Ketterle \cite{vanDruten97}. In one dimension the interaction
between bosons plays an essential role due to the strong constraint in phase
space\cite{Solyom79}. The question of BEC in a quasi one-dimensional system is
therefore more complicated.  The purpose of this paper is to demonstrate that
under suitable experimental conditions the low energy excitations of this
system are described by a Luttinger liquid (LL) \cite{Haldane81} model. The
superfluid correlations of a LL decay algebraically and the system is not Bose
condensed. At much lower temperatures which are determined by the extension of
the trap in the longitudinal direction the spectrum of the phase fluctuations
is again cut off by finite size effects and the bosons could condense again.

The realization of a Luttinger liquid in a one-dimensional Bose gas
would be a highly non-trivial example of an interacting quantum
liquid. Fermionic systems which are believed to be described by a
Luttinger liquid include quasi one-dimensional organic
metals\cite{Gruener93}, magnetic chain compounds,quantum wires and
edge states in the Quantum Hall Effect. While these systems are always
embedded in a three-dimensional matrix and thus show a crossover to a
three-dimensional behavior at low temperatures, the trapped
one-dimensional Bose gas would provide a clean testing ground for the
concept of a Luttinger liquid.

The paper is organized as follows: First we discuss the circumstances under
which a trapped Bose gas can be considered as a one-dimensional quantum
system. Next we demonstrate in an explicit calculation that there is a gapless
mode with a linear dispersion.  We show that the Hamiltonian of the low-lying
excitations can be identified as that of a Luttinger liquid and therefore the
density-density correlation function decays algebraically. In the reminder of
the paper we discuss the implication of the algebraic decay of the
particle-particle correlation function for BEC and review the properties of a
Luttinger liquid.

We consider the Bose gas in a cylindrical symmetric trap confined to
the $z$-axis by a tight trapping potential in the $xy$-plane. If the
extension $L$ of the trap in $z$-direction is much larger than its
radius $R$, it is justified to approximate the potential in the
longitudinal direction by zero.  One-dimensional physics will be
dominant, if the temperature is much lower than the energy of the
lowest radial excitation. The energy scale is set by $\hbar
\omega_\perp$, with $\omega_\perp$ being the trap
frequency\cite{Stringari96,Griffin97}. Thus the condition for
one-dimensionality is
\begin{equation}
  \label{1Dcondition}
  \hbar \omega_\perp \gg k_{\rm B} T \; ,
\end{equation}
where $T$ the temperature of the Bose gas. A typical value for $\omega_\perp$
which has been realized in the experiments performed at the MIT by the
Ketterle group\cite{MIT97} is $2\pi \cdot 240 Hz$. In order to realize a
one-dimensional Bose gas for this value of $\omega_\perp$ the temperature has
to be lower than $1.8 nK$.  Another possibility is to increase the value
$\omega_\perp$ which might be more feasible experimentally. For instance
permanent magnets can be used to increase trap frequencies by more than an
order of magnitude\cite{Meschede97}.

Assuming that this condition for $\omega_\perp$ can be met experimentally,
we can model the system by the following Hamiltonian:
\begin{eqnarray}
   \label{Hcylindrical}
 H &=& \int d^3r \; \psi^{\dagger}(\vec r \,) \left( -\frac{\hbar^2}{2m} \Delta
                   + U(\vec r \,)  
                  - \mu \right) \psi(\vec r \,) \\
&+& \frac{1}{2}  \int d^3r \; d^3r\,' \; 
     \psi^{\dagger}(\vec r \,) \; \psi^{\dagger}(\vec r\,') \;
     g\delta(\vec r - \vec r\,')\; \psi(\vec r\,') \; \psi(\vec r \,) \; ,\nonumber 
\end{eqnarray} 
where $m$ is the atomic mass, $\mu$ is the chemical potential fixed by 
the particle number $N = \int d^3r |\psi(\vec r)|^2 $ and
$g = 4\pi \hbar^2 a/m$ is the coupling constant, with $a$ being the 
s-wave scattering length. We only consider repulsive interactions.
$U = \frac{1}{2} m \omega_{\perp}^2 (x^2 + y^2) $ is the trapping potential. 
The field operators $\psi^{\dagger}(\vec r \,)$ and $\psi(\vec r\,)$ are 
bosonic creation and destruction operators.

We now illustrate that a gapless mode for a Hamiltonian like
(\ref{Hcylindrical}) exists\cite{Haldane81,Bogoliubov47}. The
dynamics of $\psi(\vec r,t)$ is governed by the equation of motion
\begin{equation}
   \label{GPE}
   i \hbar \partial_t \psi = - \frac{\hbar^2}{2m} \Delta \psi 
                           + (U - \mu) \psi + g \psi^{\dagger} \psi \psi \; ,
\end{equation}
a one-dimensional non-linear Schr\"odinger equation, Eq. (\ref{GPE}) which
well understood\cite{Haldane81}. We merely illustrate in the following its
application to the problem of trapped bosons.  For a macroscopically occupied
ground state, the operator $\psi$ can be considered as a classical complex
field. Then Eq. (\ref{GPE}) becomes the Gross-Pitaevskii equation.  We
describe the complex field $\psi(\vec r,t)$ by its density-phase
representation: $\psi(\vec r,t) = \sqrt{\rho(\vec r,t)}\exp(i\theta(\vec
r,t))$.  A saddle point solution to Eq.  (\ref{GPE}) is given by a constant
phase and static density $\rho(\vec r,t) = \rho_0(r)$ which only depends on
the radius $r$, due to the axial symmetry of the problem. The solution of the
Gross-Pitaevskii equation in a cylindrical trap and its fluctuations in the
Thomas-Fermi approximation has been discussed in detail by E. Zaremba
\cite{Zaremba97}. We only repeat the steps of the calculation necessary for
our arguments. Expanding in small fluctuations of the phase, $\delta\theta$,
and density, $\delta\rho$, around the saddle point solution:
$$ \psi(\vec r,t) = \sqrt{\rho_0 + \delta\!\rho(\vec r, t)} 
   \; e^{i[\theta_0 + \delta \theta(\vec r,t)]} \; ,$$
we obtain the linearized equations of motion for $\delta\!\rho$ and 
$\delta \theta$ 
\begin{eqnarray}
   \hbar \, \partial_t \delta \theta &=& g \delta\!\rho 
                             - \frac{\hbar^2}{4m} \, \frac{1}{\rho_0} \, 
                               \nabla \left( \rho_0 \nabla \, 
                               \frac{\delta\!\rho}{\rho_0} \right) \; ,\\
   \hbar \, \partial_t \delta\!\rho &=& \frac{\hbar^2}{m} \, 
         \nabla \left( \rho_0 \nabla \delta \theta \right) \; \; . 
\end{eqnarray}
These equations possess a trivial solution ($\delta\!\rho = 0$ , $\delta
\theta = const.$) whose energy vanishes.  This is the Goldstone mode
corresponding to global rotations of the condensate's phase.  Radial
fluctuations can be ignored, because their energy scale is set by $\hbar
\omega_{\perp}$, the trap frequency. Thus it is justified to consider the
one-dimensional limit where the equations simplify to
\begin{eqnarray}
   \hbar \, \partial_t \delta \theta &=& g \delta\!\rho 
                             - \frac{\hbar^2}{4m} \, \frac{1}{\rho_0} \, 
                               \partial^2_z \delta\!\rho \;\; ,\\
   \hbar \, \partial_t \delta\!\rho &=& \frac{\hbar^2}{m} \rho_0 
            \partial_z^2 \delta \theta \; \; . 
\end{eqnarray}
The solutions are plane waves ($\delta\!\rho \,,\, \delta \theta 
\propto e^{i(qz-\omega t)}$) with frequencies
\begin{eqnarray}
   \label{phonon-disp}
   \hbar^2\omega^2 = \hbar ^2 v_s^2 q^2 
                   + \left(\frac{\hbar^2 }{2m} \right)^2 q^4 \;\;, \\
q = \frac{2\pi}{L}n \;\; , \;\; n=0,1,...,L-1 \nonumber
\end{eqnarray}
where the sound velocity is given by $v_s = \sqrt{g\rho_0/m}$ 
which is the Bogoliubov value for a homogeneous Bose gas. It has 
been observed experimentally in anisotropic 3D traps\cite{Andrews97}.
\begin{figure}[t]
  \protect\centerline{\epsfig{file=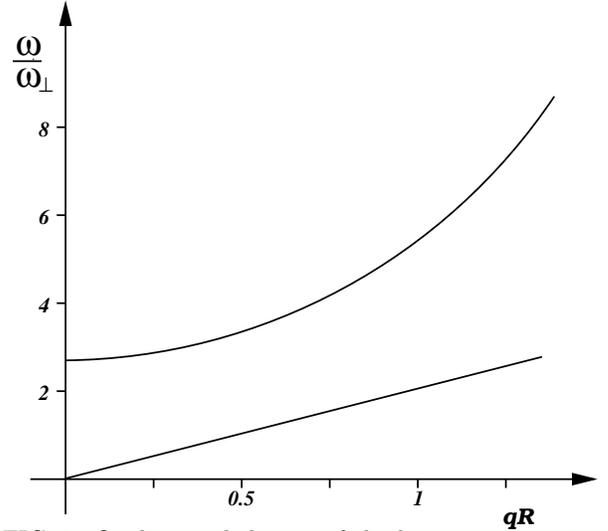,height=7cm}}
  \protect\caption{Qualitative behavior of the low energy excitation spectrum
    (lower curve) for a one-dimensional trapped Bose gas. The first radial
    excitation (upper curve) with energy $\omega\sim\omega_\perp$ is also
    shown. The discussion in the text focusses on the
    role of the lower branch.}
  \label{fig:reduction}
\end{figure}

We draw two important conclusions from this relation. The $q^2$-term cannot be
treated as a small perturbation on the energy of a non-interacting Bose gas,
hence the smallest interaction changes the excitation spectrum fundamentally.
The existence of a collective mode with linear dispersion for small $q$ is a
direct consequence of the interaction between particles. Only for a vanishing
coupling constant $g$, the spectrum reduces to that of free particles,
regardless of the ground state occupation. In a one dimensional trap ($L \gg
R$) the phase-fluctuations of the boson wave function destroy superfluid order
due to phase space constraints\cite{Hohenberg67}.  The finite size gap in
three dimensional traps($L \approx R$) introduces a cut-off in the phase space
integrals, the phase space argument does not apply
\cite{Stringari96,HoMa97} and BEC is possible.  As all trapped Bose gases are
of finite size, in principle the phonon spectrum remains discrete.  The level
splitting is only relevant in the limit
\begin{equation}
   \label{contcond}
   k_B T \ll \hbar v_s \frac{2\pi}{L} \;\; .
\end{equation}
For the MIT trap \cite{MIT97} with the length $L=0.5\,mm$, this temperature
turns out to be roughly $10^{-13} K$ (assuming $Na$ atoms). Only in this limit
the system can be in a Bose condensed phase.  If the length $L$ is not
macroscopic the gap in the lowest mode will be appreciable and there is
Bose-Einstein condensation for a finite number of particles as pointed out by
Ho and Ma \cite{HoMa97}. We stress that this is due to the smallness of $L$
and not a generic feature of the system. A setup with $L$ comparable to $R$ is
really three-dimensional.  One can check that the gap energy for a
one-dimensional trap found by Ho and Ma scales with the inverse axial
extension of the system and hence disappears for large systems. We conclude
that for $L$ satisfying condition (\ref{contcond}) there is a gapless sound
mode which inhibits the formation of a condensate at all finite temperatures.
Nonetheless the decay of coherence is only weak.  This is due to the fact that
the system can be described as a Luttinger liquid as will be shown now.

With the same approximation as for the equations of motion, the Hamiltonian in
the long-wavelength limit is:
\begin{equation}
   \label{LuttingerH}
   H = \int dz \left[ \frac{\hbar^2\rho}{2m} ({\partial_z \delta \theta})^2 \ +
       \frac{\kappa}{2\rho^2} {\delta\!\rho}^2 \right] \;,
\end{equation}
where $\rho$ is the number of particles per unit length and $\kappa$ is the 
compressibility.

The Hamiltonian, Eq. (\ref{LuttingerH}), is known as the Luttinger liquid
Hamiltonian\cite{Haldane81,Haldane79}. This concept has been mostly used to
investigate the properties of fermionic systems in one dimension. The
Luttinger liquid Hamiltonian, Eq.  (\ref{LuttingerH}), can be diagonalized by
a Bogoliubov transformation in terms of new bosonic creation and destruction
operators $b^\dagger_q , b_q$ for the long-wavelength density-fluctuation
modes.  This is possible due to the linear dispersion relation. The
Bogoliubov-transformation is given by:
\begin{eqnarray}
   \label{drho-rep}
   \delta\!\rho &=& \frac{1}{\sqrt 2} \sum_{q\ne0} e^{iqz} f_q
   \left( b_q^{\dagger} + b_{-q}^{\phantom{\dagger}} \right)\;\; ,\\
   \label{phi-rep}
   \partial_z \delta \theta &=& \frac{1}{\sqrt 2} \sum_{q\ne0} e^{iqz} g_q 
          \left( b_q^{\dagger} - b_{-q}^{\phantom{\dagger}} \right) \; .
\end{eqnarray}
The $b^{\dagger}_q , b_q$ satisfy the usual boson commutation relation
$[b_q^{\phantom{\dagger}}, b^{\dagger}_{q'}] = \delta _{q,q'}$ and $\delta
\theta$ and $\delta\!\rho$ form a pair of conjugate operators:
\begin{equation}
   \left[ \delta \theta(z) , \delta\!\rho(z') \right] = i\delta (z-z') \; .
\end{equation}
This condition fixes the functions $f_q$ and $g_q$:
\begin{eqnarray}
   f_q &=& \sqrt{|q|} e^{\alpha_q} \;, \\
   g_q &=& sgn(q) \sqrt{|q|} e^{-\alpha_q} \; ,
\end{eqnarray}
where $\alpha_q$ is the parameter of the Bogoliubov-Transformation.
Inserting the representations (\ref{drho-rep}) and (\ref{phi-rep}) for 
$\delta\!\rho$ and $\delta \theta$ respectively in the Hamiltonian 
for the fluctuations the LL in terms of the new bosonic operators is given by:
\begin{equation}
   \label{LLH}
   H = \sum_{q\ne 0} \; \hbar \omega_q \;
   \left( b_q^{\dagger} b_q^{\phantom{\dagger}} + \frac{1}{2} \right) \;\; ,
\end{equation}
with the choice $\exp(2\alpha_q) = \hbar\sqrt{\rho/mg}$. The phonon frequency
is given by $\omega_q = v_s\;|q|$ where $v_s$ is the sound velocity.

One of the striking features of a Luttinger liquid is that the model has only
two microscopic parameters, the sound velocity $v_s$ and the compressibility
$\kappa$. Another important property is that the correlation functions of the
original boson operators decay algebraically in a Luttinger liquid. The
asymptotic behavior for large distances, $z\rightarrow\infty$, of the
boson-boson and density-density correlation function is given
by\cite{Haldane81}:
\begin{eqnarray}
  \label{corrfnc}
  \langle \, \Psi^\dagger(z) \, \Psi (0) \, \rangle \; &\sim&  \;
              1 / z ^{1/\eta}  \; ,\\
  \langle \rho(z) \rho(0) \rangle - \langle \rho \rangle^2 \; &\sim& \;
          \eta / {z^2} \; ,
\end{eqnarray}
where $\eta$ is the correlation exponent. A useful naive estimate for $\eta$,
assuming that the compressibility $\kappa\sim g\rho^2$ is:
\begin{equation}
  \label{eta-estimate}
  \eta = \pi l_{B} \sqrt{\frac{2\rho}{a}}
\end{equation}
where $l_{B}=\sqrt{\hbar/\omega_\perp m}$ is the magnetic length of the trap
perpendicular to the $z$-axis and $a$ is the scattering length of the trapped
atoms. Because the interaction is weak we do not expect the exponent $\eta$ to
be renormalized substantially. For current traps the exponent $\eta$ is of the
order $\eta \sim 1000$ demonstrating that the phase coherence of the bosons
decays only very weakly and is experimentally undistinguishable from true BEC
\cite{Penrose56}.  However for steeper magnetic traps, $\omega_\perp \sim
50\;{\rm kHz}$, particle densities of $\rho \sim 10^4\; {\rm particles/cm}$
and assuming a scattering length of 110 $a_B$ for Rb\cite{Gardner95}, the
exponent $\eta$ is $\eta \sim 4$ and it should be possible to observe LL
behavior.  Below $T\sim 0.4\;{\rm nK}$ only the linear mode is excited and the
physics is described by LL physics. At still lower temperatures, $T \sim
10^{-12}$ K the finite size gap comes into play\cite{HoMa97}.

Next we compare our results to the ``two-step condensation'' picture put
forward by van Druten and Ketterle\cite{vanDruten97}. The authors consider an
{\em ideal} Bose gas in a highly anisotropic trap.  In the non-interacting
system there is no fundamental difference between the one and
three dimensions except in the density of states.  As soon as
interactions have to be considered the situation changes drastically.
Basically we have developed a more precise physical picture of the regime
which van Druten and Ketterle call the ``two-step BEC''\cite{vanDruten97}.
Our claim is that in this regime the ground state is described by a Luttinger
liquid and not by an ideal Bose gas.

Since the Luttinger liquid model has a harmonic Hamiltonian, Eq. (\ref{LLH}),
for the phase and density fluctuations, any expectation value and dynamical
correlation function of the boson operators in the long-wavelength limit can
be evaluated. Luttinger liquids are well understood and many results can be
carried over to the one-dimensional trapped Bose gas.  At larger densities the
parameters of the Luttinger liquid model will be renormalized from the saddle
point values by short range fluctuations and also by the three-dimensional
density profile of the trapped Bose gas.  The renormalized parameters can be
obtained by considering more realistic interactions in one dimension. For a
repulsive delta-function potential the sound velocity and the compressibility
have been obtained exactly\cite{Lieb63}. We are currently working on models
with longer range interactions which can be treated by the
Density--Matrix--Renormalization--Group method.  Results of this work will be
presented elsewhere\cite{Kuehner97}. In a complementary approach, we calculate
the finite size effects in the experimental setup on the dynamics of the
bosons\cite{Linn97}. Another interesting problem which we are currently
investigating is the response of the system to an impurity atom. The finite
mass leads to an unusual behavior of the mobility \cite{Kane92,Castro-Neto96}.
Also the transport properties should differ significantly from the
conventional Bose condensate if the Bose gas is in the Luttinger liquid
regime.

To summarize, we have shown under which experimental conditions a trapped
quasi-one dimensional system of interacting Bosons is described by a Luttinger
liquid Hamiltonian.  An experimental realization of such a system would
provide a clean laboratory for testing the properties of a Luttinger liquid.
Its behavior deviates significantly from the noninteracting Bose gas. Unlike
other systems which are realizations of a Luttinger liquid, three-dimensional
effects become less important for lower temperatures.  Moreover, it would be
possible to tune important parameters like the density and the length, i.e.
the trap frequency $\omega_\perp$, which is impossible in a solid.

We would like to acknowledge useful discussions with V.~Gomer, T.-L.~Ho,
W.~Ketterle, M.~Ma, D.~Meschede, A. A. Nerseseyan, A.~J.~Millis, V.~Rittenberg
and H.-J.~Schulz. H. M.  acknowledges the hospitality of the Aspen Center for
Physics.

\end{document}